# Selection of Robust Digital Communication Techniques for the Vehicle to Vehicle Communication

Subrahmanya Gunaga
B. E. Final Year : School of Electronics and
Communication Engineering
KLE Technological University
Hubballi, India

Rahul M. S.
B. E. Final Year : School of Electronics and
Communication Engineering
KLE Technological University
Hubballi, India

Varad Vinod Prabhu
B. E. Final Year : School of Electronics and
Communication Engineering
KLE Technological University
Hubballi, India

Akash Kulkarni
Research Assistant Professor : School of Electronics and
Communication Engineering
KLE Technological University
Hubballi, India

Nalini C. Iyer
Professor : School of Electronics and
Communication Engineering
KLE Technological University
Hubballi, India

*Abstract*—V2V, Vehicle to Vehicle communication has become one of the key features in achieving complete autonomy for self-driving vehicles. As digital communication forms the backbone of a vehicle to vehicle communication, we have worked on its primary building blocks: source coding, error correction and detection, and channel coding. Choosing optimal techniques for each block plays a significant role in the performance of the entire communication system. Thus, we have explored Five Source coding techniques, three Error control, and Channel coding techniques, respectively. The messages considered are categorized into three priority levels. The messages and combinations of techniques of each block were evaluated based on different comparison parameters. Based on the obtained result, the robust methods were chosen for our application.

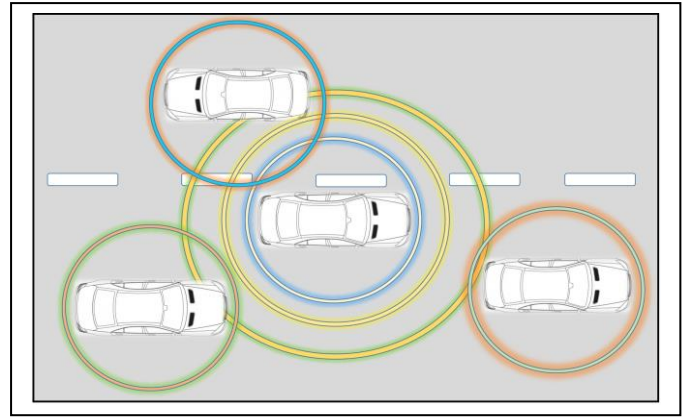

Fig. 1.Vehicle to Vehicle Communication

## I. INTRODUCTION

In this paper, we discuss the digital communication system for a Vehicle to Vehicle Communication using Dedicated Short Range Communication (DSRC) protocol.

The autonomous vehicle is one of the current trending technology. Significant work is being carried in this field to achieve complete autonomy. But, there are various hurdles to be overcome before complete autonomy can be achieved. V2V Communication proves to be a valuable asset in achieving the goal. V2V communication refers to the exchange of information between vehicles which helps in taking precise decisions. Figure 1 shows the short-range vehicle to vehicle communication.

There are different ways in which the data can be transferred between vehicles, but one of the prominent methods is through LTE-V2X (5G technology). 5G refers to fifth-generation technology for cellular networks. Vehicles use 5G signals to transmit the messages. The reason for using 5G is a high data rate and low latency. Another type of protocol for establishing communication is DSRC, which is the abbreviation for Dedicated Short Range Communication. The frequency of operation of DSRC is in a 5.9 GHz band. Due to high frequency, the maximum range of the signal is not high but it is enough for the application. For the vehicle to vehicle communication DSRC finds its applications as it has short-range and used for peer-to-peer communication. The backbone of these protocols is the principle of a digital communication system. At the root level, any digital communication system has three major parts, they are, source coding, error detection and correction, and channel coding. There are various techniques in three major parts of the digital communication system mentioned. The selection of the optimal technique for each of the techniques based on the application will help to design an efficient communicating system.

Here, we explore the work that has been done in various sub-systems of the digital communication system. The building block of the digital communication system is source coding, and decoding, error correction coding, and decoding, channel coding, and decoding, modulation, and demodulation techniques as shown in Figure 2.

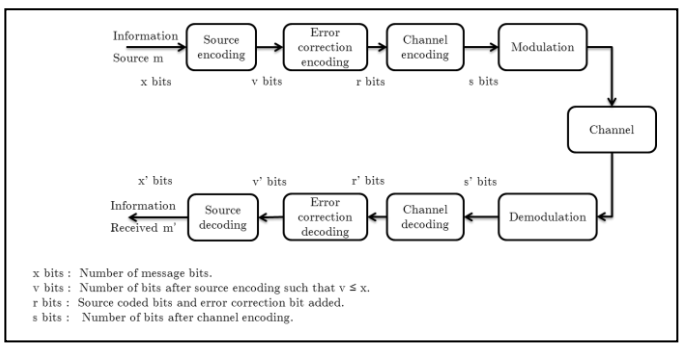

Fig. 2. Block diagram of digital communication system

Several works have been done by researchers towards source coding for text data. Authors of [1] propose a novel entropy-based coding method for data formats such as text, image, audio, and video. The proposed method takes into account the frequency of symbols in a sequence. The symbols are assigned a particular group and coded using the number of bits corresponding to that group based on the frequency of symbols. In the paper [2], the author concludes from the experimental results that the double minimum variance Huffman coding is more efficient and has better compression than the single Huffman coding at the cost of increased computational and space complexity. The authors of the paper [3] discuss various lossless compression methods and explain the advantages, disadvantages, principles of operation, and features of each of these algorithms. The paper compares the compression ratio, CPU time, memory cost, and size of the executable code. LZ77, PPM, and BWT algorithms are suitable for electric vehicle communication.

Several, works have been done to detect and correct single-bit errors, random error, and burst error that occur in the received bits. The author in the paper [4] compares Reed-Muller codes with other error correction and detection codes for multiple bit errors. Reed-Muller codes are suitable for long-range communication due to its ability to correct multiple errors. Authors in the paper [5] have proposed a coding technique, known as double error correction (DEC). The decoding complexity is slightly high but is capable of reducing parity bits. Schemes related to Orthogonal Latin Square (OLS) codes have a greater number of parity bits compared to this method. It also has a lesser delay in decoding when compared to that of Bose Chaudhuri Hocquenghem (BCH) codes. In the paper [6] the coding scheme called Data negation codes is proposed. Data negation codes are used for single or multiple random errors and in its modified form can be used to correct burst errors. The efficiency of the product codes using data negation is less than that of the data negation codes, but are capable of correcting multiple random errors as well as burst errors.

Channel Coding is one of the major components of the digital communication protocol. It plays a significant in correcting and detecting errors caused by noise interference when travelling through channel. There have been works in this field for finding the best coding technique, some of them are mentioned here, Bashar Tahir *et al.* [7], has provided the significance of channel coding in a communication system. There is information regarding the requirement of different channel coding techniques based on the application being used. The comparison between convolutional, polar, turbo, and LDPC codes has been provided. And the parameters considered for the comparison are BER, the number of iterations is used for comparison. It is observed that LDPC provides better performance for a higher coding rate. Turbo and Polar are a good option when we are dealing with a lower length of code. Walled Abdulwhab *et al.* [8], has presented a need for channel coding in the communication system. The main emphasis is on the use of LDPC or Polar for 5G. LDPC codes are of commercial importance as they are implemented in real systems. These are efficient in terms of energy usage and area. This is best suited for a higher coding rate and a greater code length. Convolutional code has the least performance when compared to that of LDPC, Polar, and Turbo. There are two shortcomings of LDPC. Complexity increases at a low rate, and also there is a degradation of performance for more than a dB. Othman O. Khalifa *et al.* [9], addresses the performance of Channel Coding when using Viterbi Algorithm and LDPC codes which BPSK modulated and the noise factor is AWGN. When the block length is very large the performance of the LDPC code is comparatively better than the Turbo code. For decoding these are less computationally intensive. Another advantage is the parallelization capability helps hardware implementation of LDPC codes. The benefits of Viterbi decoding is it has got a fixed decoding time. Hence, suitable for hardware decoder completion. Diego Lentner *et al.* [10] has provided the comparison between Turbo, Polar, binary and non-binary LDPC and tail-biting convolutional code. The parameter considered is a bit error rate parameter. The bit-error performance gradually degrades as there is an increase in message length. For the simulation, the code rate is set to 1/2 and the message length, k is considered to be less than 512. Tail biting convolutional code outperforms LTE. Turbo code for k $\leq$ 128. The takeaway is that the performance of Tail-biting convolutional codes with medium size memory elements is commendable for very short message lengths (64 bits), and have decoders with lower complexity.

II. PROPOSED METHODOLOGY

The digital communication system can be expressed in terms of three main stages which are shown in Figure 3. Work has been done in respective fields to find out the robust technique for the V2V application.

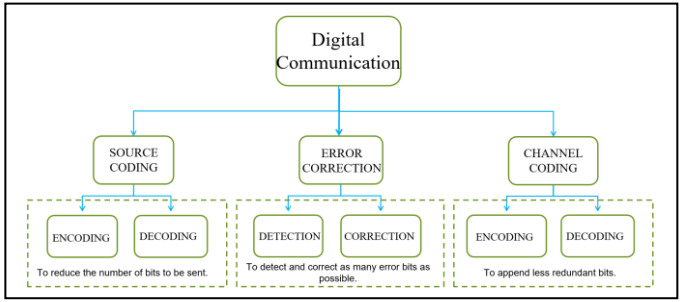

Fig. 3. Prominent stages of digital communication system

## A. Source Coding

There are many lossless source coding techniques. In this paper, we compare the performance of compression of Huffman coding, Arithmetic coding, Lempel-Ziv-Welch coding with application-specific approaches like Abbreviation based coding, and Probability-based coding for the communication of safety messages for the vehicle to vehicle communication. For comparison of algorithms, the $20$ safety messages for the vehicle to vehicle communication is considered as shown in Table I.

### 1) Huffman coding:

Huffman coding [11] is a lossless source coding technique. It is an entropy-based technique that is used to generate a variable-length code for each of the source symbols. It works on the principle that more frequent the symbols occur then it is represented by fewer bits. On the other hand, the less frequent symbols are mapped to more number of bits.

For the transmission of safety messages for the vehicle to vehicle communication. Based on the probability of occurrences of small case letters, and space in the English language the Huffman tree is generated. The advantage of using a fixed tree is that the overhead of the generation of the tree is reduced, transmitting the mapping of the symbols and their respective code-word to the receiver is reduced. There are adaptive versions of Huffman coding but the disadvantages associated with them are they are prone to errors, the compression achieved is not so significant, and each time for a new symbol the tree has to be updated which is computationally expensive for this application.

### 2) Arithmetic coding:

Arithmetic coding [12] is an entropy-based coding used for lossless data compression. Arithmetic coding maps the entire message into a single number. For the long sequences with skewed distribution and a small number of symbols, arithmetic coding performs better than Huffman coding.

For transmitting the messages for the vehicle to vehicle communication based on the distribution of small case letters and space in the English language the message is mapped to its corresponding tag using the implementation mentioned in [12]. The received tag is decoded at the receiver.

### 3) Lempel-Ziv-Welch (LZW) coding:

Lempel-Ziv-Welch (LZW) [13] is dictionary-based coding. It is a form of universal lossless source coding technique. Lempel-Ziv-Welch algorithm performance is better if there is a repetition of the pattern in the message to be transmitted.

In our implementation, we have initialized the dictionary to contain small case English alphabets and space. In the original implementation as mentioned in [13], the sequence of 8-bit data is mapped to a fixed-length 12-bit code. A small modification is done instead of representing each sequence of 8-bit data as a fixed-length 12-bit code, the first four bits are appended at the starting of the compressed sequence of bits to indicate the fixed number of bits used to represent each code. This modification is helpful as it reduces the number of bits to be transmitted.

### 4) Application-specific source coding:

In this sub-section, we explore the application-specific methods for source coding of a vehicle to vehicle safety messages communication. We have identified 20 messages which are used for vehicle to vehicle communication as shown in Table I. Here instead of source coding for generalized purposes, we consider the common messages for a vehicle to vehicle communication and perform source coding of these messages. These messages and their corresponding mapping are known to both the transmitter and the receiver. Thus, using these application-specific source techniques can be helpful to reduce the number of bits to be transmitted. Thus, as the number of bits to be transmitted is reduced the communication can be faster.

#### a) Abbreviation-based coding:

In this method of coding similar to Ham radio abbreviations, the safety messages are assigned abbreviations as shown in the third column of Table I. Instead of transmitting the entire safety message, the abbreviation corresponding to the message is transmitted. Thus the number of bits to be transmitted is reduced. As shown in Table I any safety message is represented by a fixed three-character abbreviation. Thus any safety message can be transmitted using 24 bits.

#### b) Probability-based coding:

The principle used behind probability-based coding is that instead of coding each symbol of the message to be transmitted, the entire message is mapped to a code-word. Since priority-based or emergency based messages need to be transmitted quickly they need to be represented with a fewer number of bits as possible.

Since the number of occurrences of these messages is not known. To get the probability of occurrences of each message, the messages are classified into three priority levels $P_1$, $P_2$, and $P_3$ respectively as shown in the fourth column of Table I. The messages with priority $P_1$ are the messages which occur frequently or are emergency messages and should be represented with a fewer number of bits. Similarly, the message with priority $P_2$ are the messages which occur less frequently or not emergency message and the messages with priority $P_3$ are the messages which are used for additional precautions or additional information thus it can be represented by comparably more number of bits. Based on this priority levels the probability of occurrence of each message is assigned as shown in the fifth column of Table I. Using Huffman coding based on the probability of occurrence of messages the code-word for each message is obtained as shown in the sixth column of Table I.

*B. Error Correction and Detection*

*1) Hamming Code:*

This technique was developed by R. W. Hamming. Hamming code belongs to linear error correction coding. Only 3 redundancy bits are required for a block with 4 data bits called (7, 4) Hamming code.

In this technique, the redundancy bits are added at the position of $2^n$ th position. The redundancy bits are dependent on even or odd parity. The number of redundancy bits calculated using $2^r \geq (m+r+2)$, where r represents the number of redundant bits, and m represents the number of input data bits.

In the case of even parity, If the number of 1's is odd then the parity bit value is set to 1. If the total number of 1's is even, the parity bit's value is 0.

In the case of odd parity, If the number of 1's is even then the parity bit value is set to 1. If the total number of 1's is odd, the parity bit's value is 0.

To find redundancy bits, the position number is written in binary digits. The bits at the power of 2 are considered as the redundancy bits. The first redundant bit is obtained by considering the number of 1's at the location of the bits which has a 1 at the least significant bit in its binary representation like bits with positions 1, 3, 5, 7, 9. On a similar basis, the second redundant bit is obtained by considering the number of occurrences of 1's at the bit location which has a 1 at the second bit from the least significant bit in its binary representation like bits with positions 2, 3, 6, 7. Similarly, the redundancy bit at location 4 is obtained by considering all the occurrences of 1's at the bit location which has a 1 at the third bit from the least significant bit in its binary representation like bits with position 4, 5, 6, 7. The encoded code-word (data bits + redundancy bits) is transmitted by the transmitter through the channel. The error has occurred due to noisy channels or distraction in channels. At the receiver side, the error position is identified using even or odd parity. A detected error is recovered by complimenting the bit. Hamming code helps to find single-bit errors.

TABLE I: BASIC SAFETY MESSAGES FOR VEHICLE TO VEHICLE COMMUNICATION

| S. L. No. | Safety Messages | Abbreviations | Priority | Probability | Probability-based code |
|---|---|---|---|---|---|
| 1 | left turn ahead | LTA | $P_2$ | 50/1075 | 00111 |
| 2 | right turn ahead | RTA | $P_2$ | 50/1075 | 00110 |
| 3 | emergency ahead | EGA | $P_1$ | 100/1075 | 101 |
| 4 | emergency braking | EGB | $P_1$ | 100/1075 | 100 |
| 5 | brakes applied | BKA | $P_1$ | 100/1075 | 111 |
| 6 | lane change alert | LCA | $P_2$ | 50/1075 | 00001 |
| 7 | queue warning | QEW | $P_3$ | 25/1075 | 001001 |
| 8 | hump warning | HMW | $P_3$ | 25/1075 | 001000 |
| 9 | pedestrian crossing ahead | PCA | $P_1$ | 100/1075 | 110 |
| 10 | work in progress ahead | WPA | $P_3$ | 25/1075 | 001011 |
| 11 | leave way for the ambulance | LWA | $P_1$ | 100/1075 | 011 |
| 12 | intersection ahead | ISA | $P_2$ | 50/1075 | 00000 |
| 13 | taking left turn | TLT | $P_2$ | 50/1075 | 00011 |
| 14 | taking right turn | TRT | $P_2$ | 50/1075 | 00010 |
| 15 | road condition not good | RNG | $P_3$ | 25/1075 | 001010 |
| 16 | allow overtake | AWO | $P_3$ | 25/1075 | 010101 |
| 17 | allowed overtake | AEO | $P_3$ | 25/1075 | 010100 |
| 18 | searching for parking | SFP | $P_3$ | 25/1075 | 01011 |
| 19 | taking u turn | TUT | $P_2$ | 50/1075 | 01001 |
| 20 | vehicle turning in front | VTF | $P_2$ | 50/1075 | 01000 |

*2) Tornado Code:*

Tornado code belongs to erasure coding that supports error correction. Tornado code helps to correct the multiple bits. The tornado code structure is constructed using Gallager code based randomly connected irregular bipartite graphs. The data is divided into blocks according to block size. For our project, we constructed the structure of (12, 7) tornado code. The right side of the graph represents redundancy bits (c) and the left side represents data bits (d). The redundancy bits are calculated by modulo 2 depending on the structure of code. The redundancy bits (c) are appended to data bits (d). The encoded bits are transferred through a channel. The error may occur during transmission. The errors are identified and corrected while decoding. The erroneous data bits are decoded by performing modulo 2 operation based on the structure of the block. The data rate of this technique depends on the structure of the bipartite graph.

*3) Data Negation Code:*

Data negation belongs to the weight-based binary error correction coding technique. It helps to find random errors. The coding efficiency will be *50%* for data negation coding that is even, odd and zero weights. The redundancy bits are padded to data bits. The redundancy bits are added based on the weights that are even, odd and zero weights. The number of redundancy bits is equal to the number of data bits. If the number of 1's is zero that is zero weight, the redundancy bits are the same as data bits. If the number of 1's is even that is even weight, the redundancy bits are the same as data bits. If the number of 1's is odd that is odd weight, then the data bits are complemented to obtain the redundancy bits. For data negation coding the number of encoded bits is always double that of the number of the input data bits. Figure *4* shows the flow chart of data negation coding.

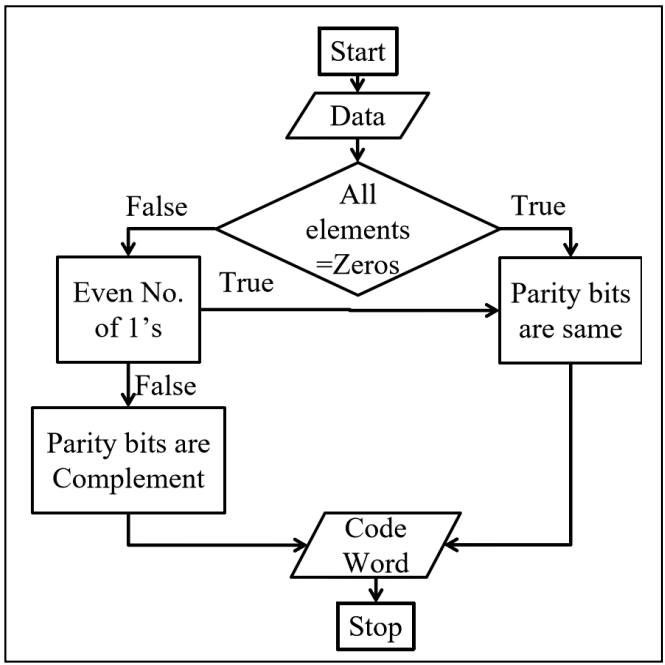

Fig. 4. Flow chart of data negation

The errors may occur to encoded bits while transmitting through noisy channels or distortion. At the receiver, erroneous data is identified by XOR the data and redundancy bits. The identified error bit is corrected by complementing that bit. The data rate will be 50%.

*C. Channel Coding*

Here we have considered three coding techniques, those are Convolutional, Turbo and, LDPC code. Here we discuss the working of these techniques.

*1) Convolutional Code:*
P.Elias was the founder of this coding technique. It is one of the most popular coding technique. Here the input message is convoluted with the designed logic and the transmitted. The convolutional operation introduces some redundant bits into the message. Blocks of data and continuous data stream, both these format can be coded using this technique. For every unit if time, k bit message are fed as an input, the output will be n bit coded block, where n > k. The coded n bit not only depends on the k input, but also on the m previous messages. Here the memory block has a significant role in coding, thus the memory of order m is taken into consideration for designing the generator sequence. Thus the parameters of convolutional code are,

n - number of output bits k - number of input bits m - number of memory registers

Working:
The encoding of the data is accomplished through shift registers and combinatorial logic that performs modulo addition. The number of output lines decides the rate. Generator sequence is required for performing convolution, and this sequence is dependent on the memory blocks and how the blocks are connected to the input lines.

Generator Sequences are referred to impulse responses.

Mathematical Equations:
Consider an Encoder which has k inputs and n outputs.
It is represented as *(n, k ,m)*. Here m is number of memory element.

Sequence of input messages at the $i^{th}$ terminal is given by Equation 1:
$$Msg^{(i)} = (Msg_0^{(i)}, Msg_1^{(i)}, Msg_2^{(i)}, Msg_3^{(i)}, Msg_4^{(i)}, .....) \quad (1)$$
$$\text{for } 1 \leq i \leq k,$$

Sequence of Output Code at the $j^{th}$ terminal is given by the Equation 2:
$$C\_word^{(j)} = (C\_word_0^{(j)}, C\_word_1^{(j)}, C\_word_2^{(j)}, C\_word_3^{(j)}, C\_word_4^{(j)}, ..................) \quad (2)$$
$$\text{for } 1 \leq j \leq n,$$

An (n, k, m) convolutional code is specified by k x n generator sequence as shown in Equation 3:
$$\begin{array}{l} Gen_1^{(1)}, Gen_1^{(2)}, Gen_1^{(3)}, ............ Gen_1^{(n)} \\ Gen_2^{(1)}, Gen_2^{(2)}, Gen_2^{(3)}, ............ Gen_2^{(n)} \\ Gen_3^{(1)}, Gen_3^{(2)}, Gen_3^{(3)}, ............ Gen_3^{(n)} \\ .... \\ .... \\ .... \\ .... \\ Gen_k^{(1)}, Gen_k^{(2)}, Gen_k^{(3)}, ............ Gen_k^{(n)} \end{array} \quad (3)$$

The output code sequence is given by Equation 4:
$$\begin{array}{l} C\_word^{(1)} = Msg^{(1)}*Gen_1^{(1)} + Msg^{(2)}*Gen_2^{(1)} + Msg^{(3)}*Gen_3^{(1)} \\ +.....+ Msg^{(k)}*Gen_k^{(1)} \\ C\_word^{(2)} = Msg^{(1)}*Gen_1^{(2)} + Msg^{(2)}*Gen_2^{(2)} + Msg^{(3)}*Gen_3^{(2)} \\ +.....+ Msg^{(k)}*Gen_k^{(2)} \\ C\_word^{(3)} = Msg^{(1)}*Gen_1^{(3)} + Msg^{(2)}*Gen_2^{(3)} + Msg^{(3)}*Gen_3^{(3)} \\ +.....+ Msg^{(k)}*Gen_k^{(3)} \\ ..... \\ ..... \\ ..... \\ ..... \quad (4) \\ C\_word^{(n)} = Msg^{(1)}*Gen_1^{(n)} + Msg^{(2)}*Gen_2^{(n)} + Msg^{(3)}*Gen_3^{(n)} \\ +.....+ Msg^{(k)}*Gen_k^{(n)} \end{array}$$

Here the convolution operation is represented by : *

For our implementation the chosen generator sequence and value of (n, k, m) are:
(n, k, m) : (2, 1, 4)

Gen1 : [ 1 0 1 1 ]
Gen2 : [ 1 1 1 1 ]

Convolutional Code can be represented in following ways:
- Generator, Trellis, Tree, and State Diagram Representation.

*2) Turbo Code:*

These are error correcting codes introduced in the year 1993. These are known for their performance with comparatively lower complexity encoding and decoding approaches.

In simple words the Turbo codes can be seen as an combination of Convolution and Turbo Code, as it performs similar operation with an interleaver in between two convolution blocks. This interleaver helps in overcoming burst error due to interference in channel.

Figure 5 shows the general block representation of Turbo coding.

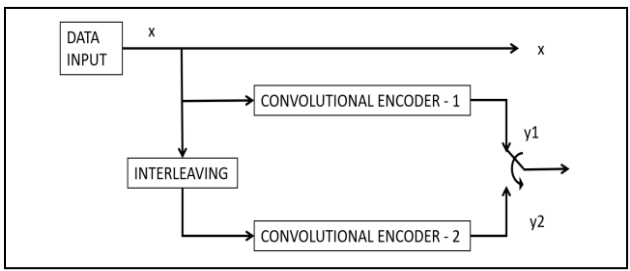

Fig. 5. Block diagram of turbo coding

In the Figure 5, x refers to the input data. This data is directly passed onto to the first Encoder -1. The Encoder - 2 receives the message after passing through the interleaver. The combination of message bit, and parity bits from the two encoders results in coded data. The main function of Interleaving is to overcome the burst errors. Burst errors are those where there is a continuous error bits in the received data.

*3) LDPC Code:*

LDPC stands for Low Density Parity Check Matrix, it is a linear block code and is defined by parity check matrix. The sparsity of the parity check matrix is the reason for this technique to be referred as low density. Sparsity is in terms of number of 1's present in the matrix.

Figure 6 presents the steps involved in encoding of message bits using LDPC. The input data U is multiplied with the generator matrix G which is obtained through parity check matrix H. Wr and Wc are the number of 1's in row and column. The generated code word is checked using the property of orthogonality. Since it is a matrix operation, the dimension of the matrix plays a significant role. There are several parameters which must be predefined, such as the dimension, Wr and Wc, and expansion factor z. Expansion factor is concerned with the number of times the column is shifted. There are two standard Base Graph dimensions, BG1 = 46 x 68 BG2 = 42 x 52.

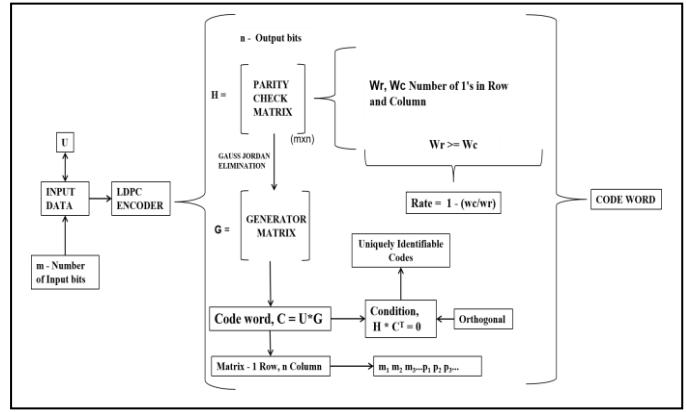

Fig. 6. Block diagram of LDPC coding

The generator matrix G, is defined using Equation 5:
$$\text{code\_word} = G^T \text{msg} \quad (5)$$
where,
- $\text{code\_word} = [\text{code\_word}_1, \text{code\_word}_2, \text{code\_word}_3, \ldots, \text{code\_word}_N]^T$ - Code-word
- $\text{msg} = [\text{msg}_1, \text{msg}_2, \text{msg}_3, \ldots, \text{msg}_k]^T$ – Message word
- G = k by n Generator matrix.

*a) Encoding:*
$$\text{code\_word} = G^T \text{msg} \quad (6)$$

We define a complete set of successful parity-checks using Equation 7:
$$H \times \text{code\_word} = 0 \quad (7)$$
where,
- $\text{code\_word} = [\text{code\_word}_1, \text{code\_word}_2, \text{code\_word}_3, \ldots, \text{code\_word}_N]^T$
- $H_{(N-K) \times N}$ = (N-K) by N parity-check matrix.

The location of the parity-bits in the code-word is arbitrary, therefore we will form our code-word using Equation 8:
$$\text{code\_word} = [\text{parity} : \text{msg}]^T \quad (8)$$
where,
- $\text{msg} = [\text{msg}_1, \text{msg}_2, \text{msg}_3, \ldots, \text{msg}_k]^T$ – Message word
- $\text{parity} = [\text{parity}_1, \text{parity}_2, \text{parity}_3, \ldots, \text{parity}_{N-k}]^T$ – Parity Bits

Therefore:
$$H[\text{parity} : \text{msg}]^T = 0 \quad (9)$$

H can be partitioned as shown in Equation 10:
$$H = [X : Y] \quad (10)$$
where,

- $X$ = (N-k) by (N-k) sub-matrix.
- $Y$ = (N-k) by k sub-matrix.

From this we can find:
X parity + Ymsg = 0     (11)

Using modulo-2 arithmetic we can solve for p using Equation 12
parity = $X^{-1}$Ymsg     (12)

Then we solve for c using Equation 13 as:
code_word = $[(X^{-1}Y)^T : I]^T$msg     (13)
Where I is the k by k identity matrix and we define G as Equation 14:
G = $[(X^{-1}Y)^T : I]$     (14)

### III. RESULTS AND DISCUSSIONS

The implementation of the algorithms was done on Windows 10, a 64-bit operating system using MATLAB Version 9.4 (R2018a). The parameters that were considered for the simulation are:
1. Channel - Additive White Gaussian Noise (AWGN) channel with $E_b/N_0$ = 2.
2. Modulation technique - 16-bit Quadrature Amplitude Modulation (QAM).

The designed system has been simulated for three different messages, whose priorities vary from 1 to 3.
- Priority 1: Emergency Ahead
- Priority 2: Left turn ahead
- Priority 3: Road condition not good

### A. Compression ratio

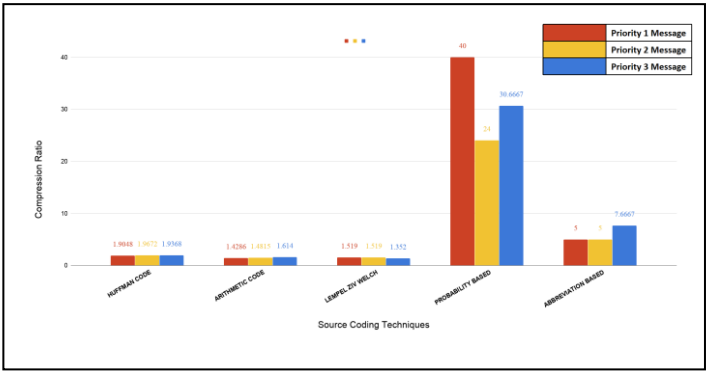

Fig. 7. Comparison of compression ratio for the three priority messages considered

Figure 7 presents a comparison between the compression ratio for all source coding techniques. Figure 7 represents the result for messages of priority 1, 2, and, 3 respectively. Probability-based code has a commendable compression ratio.

### B. Error detection and correction: Bit Error Rate

Parameter BER is computed for Error Detection and Correction codes. This comparison provides us which error control technique will perform better. Figure 8, Figure 9, and Figure 10 are the plots of BER for all the combinations. Considering all the priority messages, Tornado Code can be seen performing better than other techniques.

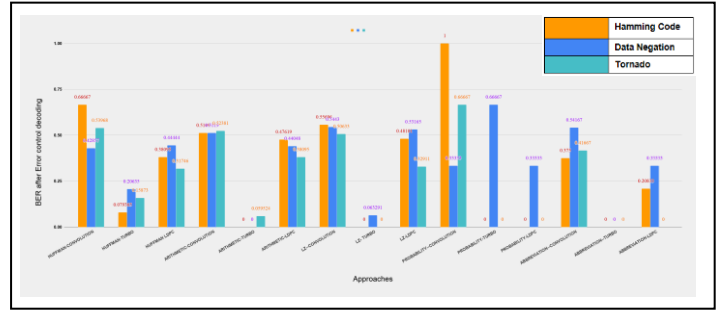

Fig. 8. BER after error correction decoding for the message Emergency ahead

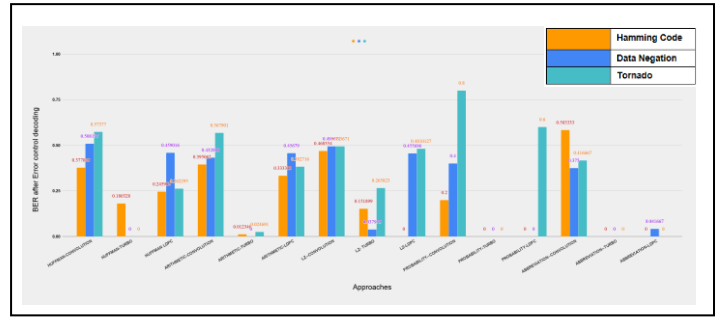

Fig. 9. BER after error correction decoding for the message Left turn ahead

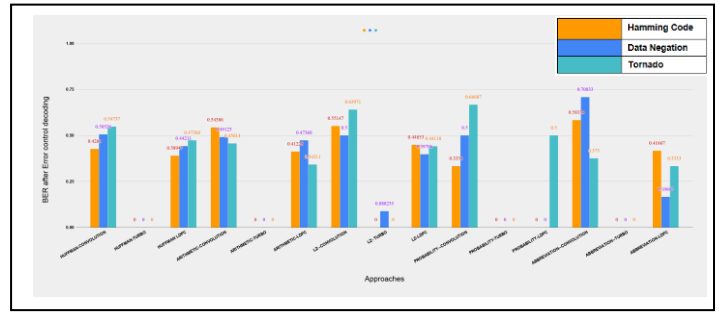

Fig. 10. BER after error correction decoding for the message Road condition not good

### C. Channel Coding: Bit Error Rate

Bit Error Rate is a parameter computed for all the 45 combinations and have been plotted for three messages which are presented in Figure 11, Figure 12, and, Figure 13. From all the three plots it can be noted the BER is lowest for those combinations which have Turbo code as the channel coding technique. There are cases where the BER is low for LDPC and Convolutional, but when all three messages are taken into account, the Turbo code performs better.

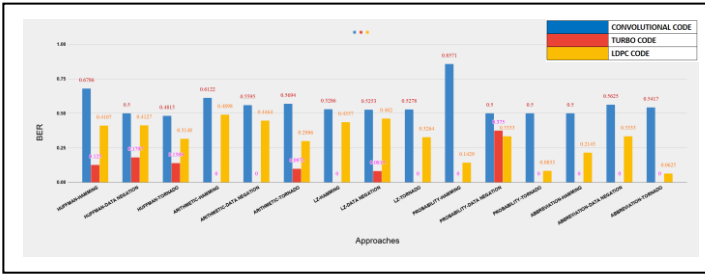

Fig. 11. BER after channel decoding for the message Emergency ahead

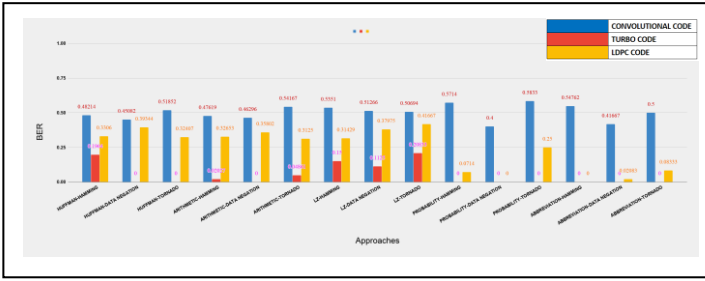

Fig. 12. BER after channel decoding for the message Left turn ahead

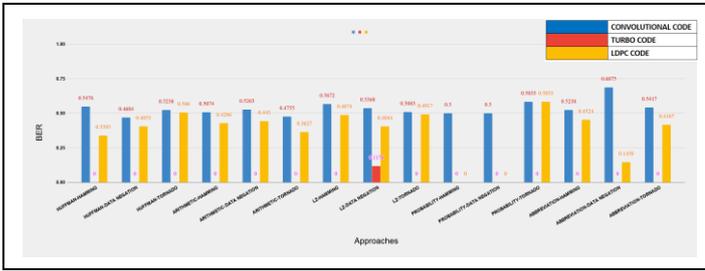

Fig. 13. BER after channel decoding for the message Road condition not good

## IV. Conclusions

Out of the 45 combinations, one combination was selected to be optimal for the designed application. The selection was made based on the performance of each technique for the mentioned parameter. Table II presents the final technique chosen from each field and the reason for the selection.

TABLE II: REASON FOR SELECTION OF THE TECHNIQUES

| Coding technique | Technique selected in the coding technique | Reason for the selection |
|---|---|---|
| Source coding | Probability-based coding | Fewer number of bits for transmission. |
| Error control coding | Tornado coding | Able to detect multiple error with smaller bit error rate. |
| Channel coding | Turbo coding | Least number of bits to be transmitted with smaller value of bit error rate. |

## References

[1] M. Ezhilarasan, P. Thambidurai, K. Praveena, S. Srinivasan, and N. Sumathi, "A new entropy encoding technique for multimedia data compression," in International Conference on Computational Intelligence and Multimedia Applications (ICCIMA 2007), vol. 4, 2007, pp. 157–161.

[2] G. S. Sandeep, B. S. S. Kumar, and D. J. Deepak, "An efficient lossless compression using double huffman minimum variance encoding technique,"in 2015 International Conference on Applied and Theoretical Computing and Communication Technology (iCATccT), 2015, pp. 534– 537.

[3] C. Shuai, S. Li, and H. Liu, "Comparison of compression algorithms on vehicle communications system," in 2015 IEEE Advanced Information Technology, Electronic and Automation Control Conference (IAEAC), 2015, pp. 91–95.

[4] J. Singh and J. Singh, "A comparative study of error detection and correction coding techniques," in 2012 Second International Conference on Advanced Computing Communication Technologies, 2012, pp. 187– 189.

[5] S. Liu, J. Li, P. Reviriego, M. Ottavi, and L. Xiao, "A double error correction code for 32-bit data words with efficent decoding," IEEE Transactions on Device and Materials Reliability, vol. 18, no. 1, pp. 125–127, 2018.

[6] N. Shribala, P. Srihari, and B. C. Jinaga, "Multiple error correction binary channel coding scheme," in 2013 International Conference on Green Computing, Communication and Conservation of Energy (ICGCE), 2013, pp. 10–16.

[7] B. Tahir, S. Schwarz, and M. Rupp, "Ber comparison between convolutional, turbo, ldpc, and polar codes," in 2017 24th International Conference on Telecommunications (ICT), 2017, pp. 1–7.

[8] W. K. Abdulwahab and A. Abdulrahman Kadhim, "Comparative study of channel coding schemes for 5g," in 2018 International Conference on Advanced Science and Engineering (ICOASE), 2018, pp. 239–243.

[9] P. Chennnapragada, "Performance analysis of 4g systems with channel coding algorithms," HELIX, vol. 8, pp. 2742–2746, 01 2018.

[10] O. Iscan, D. Lentner, and W. Xu, "A comparison of channel coding schemes for 5g short message transmission," in 2016 IEEE Globecom Workshops (GC Wkshps), 2016, pp. 1–6.

[11] D. A. Huffman, "A method for the construction of minimum-redundancy codes," Proceedings of the IRE, vol. 40, no. 9, pp. 1098–1101, 1952.

[12] K. Sayood, Introduction to Data Compression (2nd Ed.). San Francisco, CA, USA: Morgan Kaufmann Publishers Inc., 2000.

[13] Welch, "A technique for high-performance data compression," Computer, vol. 17, no. 6, pp. 8–19, 1984.

[14] R. A. Bedruz and A. R. F. Quiros, "Comparison of huffman algorithm and lempel-ziv algorithm for audio, image and text compression," in 2015 International Conference on Humanoid, Nanotechnology, Information Technology,Communication and Control, Environment and Management (HNICEM), 2015, pp. 1–6.

[15] K. Sharma and K. Gupta, "Lossless data compression techniques and their performance," in 2017 International Conference on Computing, Communication and Automation (ICCCA), 2017, pp. 256–261.

[16] S. V. Viraktamath, M. V. Koti, and M. M. Bamagod, "Performance analysis of source coding techniques," in 2017 International Conference on Computing Methodologies and Communication (ICCMC), 2017, pp. 689–692.

[17] O. Jalilian, A. T. Haghighat, and A. Rezvanian, "Evaluation of persian text based on huffman data compression," in 2009 XXII International Symposium on Information, Communication and Automation Technologies, 2009, pp. 1–5.

[18] M. Annamalai, D. Shrestha, and S. Tjuatja, "Experimental study of application specific source coding for wireless sensor networks," 11 2008.

[19] J. S. Vitter, "Design and analysis of dynamic huffman codes," J. ACM, vol. 34, no. 4, p. 825–845, Oct. 1987. [Online]. Available: https://doi.org/10.1145/31846.42227

[20] P. M. Shah, P. D. Vyavahare, and A. Jain, "Modern error correcting codes for 4g and beyond: Turbo codes and ldpc codes," in 2015 Radio and Antenna Days of the Indian Ocean (RADIO), 2015, pp. 1–2.

[21] P. Farkaˇs, T. Janvars, K. Farkaˇsov´a, and E. Ruˇzick´y, "On run-length limited error control codes constructed from binary single parity check product codes," in 2018 Cybernetics Informatics (K I), 2018, pp. 1–4.

[22] J. Wolf, A. Michelson, and A. Levesque, "On the probability of undetected error for linear block codes," IEEE Transactions on Communications, vol. 30, no. 2, pp. 317–325, 1982.

[23] F. Xiao, "Researchonldpccodedecodingalgorithmbasedonscheduling strategy," in 2016 Chinese Control and Decision Conference (CCDC), 2016, pp. 3816–3821.

[24] D. Dechene, K. Peets, and J. Cheng, "simulated performance of lowdensity parity-check codes," LAKEHEAD UNIVERSITY CANADA, Article, 2006.